\documentclass[%
 reprint,
 groupedaddress,
 showpacs,
 amsmath,amssymb,
 aps,
 prb,
floatfix,
]{revtex4-1}

\usepackage{graphicx}%
\usepackage{dcolumn}%
\usepackage{bm}%

\begin{document}

\title{Adsorption and STM imaging of TCNE on Ag(001): An ab-inito study}%

\author{Thorsten Deilmann}%
  \email{thorsten.deilmann@uni-muenster.de}%
\author{Peter Kr\"uger}%
\author{Michael Rohlfing}%
\affiliation{Institut f\"ur Festk\"orpertheorie, Westf\"alische Wilhelms-Universit\"at M\"unster, D-48149 M\"unster, Germany.}%
\author{Daniel Wegner}%
\affiliation{Physikalisches Institut and Center for Nanotechnology (CeNTech), Westf\"alische Wilhelms-Universit\"at M\"unster, D-48149 M\"unster, Germany.}%

\date{9 January 2014}%

\begin{abstract}%
We investigate the adsorption of a single tetracyanoethylene (TCNE) molecule on the silver (001) surface.
Adsorption structures, electronic properties, and scanning tunneling microscopy (STM) images
are calculated within density-functional theory.
Adsorption occurs most favorably in on-top configuration, with the C=C double bond directly
above a silver atom and the four N atoms bound to four neighboring Ag atoms.
The lowest unoccupied molecular orbital of TCNE becomes occupied due to electron transfer from the substrate.
This state dominates the electronic spectrum and the STM image at moderately negative bias.
We discuss and employ a spatial extrapolation technique for the calculation of STM and
scanning tunneling spectroscopy (STS) images.
Our calculated images are in good agreement with experimental data.
\end{abstract}

\maketitle

\section{Introduction}\label{sec:intro}
Organic molecules adsorbed on metal surfaces are in the focus of current research both from
a fundamental as well as an applied point of view.
In this context, the strong electron acceptor tetracyanoethylene (TCNE) has recently attracted a
lot of interest.
This molecule (C$_6$N$_4$) consists of a \text{C=C} double bond and four cyano groups.
Its disposition for electron-transfer reactions and 
its ability to bond to as many as four metal ions makes it a building block
for technologically interesting assemblies like molecular magnets.\cite{Miller, MolMagVTCNE} 
Transition metal compounds of TCNE show a rich magnetic behavior, including
systems with a high Curie temperature up to $400$\,K.
Thin films of these systems deposited on a substrate offer appealing
applications.
For a deeper understanding of a molecular magnet
built on a metal surface,\cite{Weg09} a good knowledge of its components and
their interaction with the substrate is essential. 
To this end, in Ref.~\onlinecite{Weg08} TCNE molecules adsorbed on noble metal surfaces have been investigated employing
scanning tunneling microscopy (STM) and scanning tunneling spectroscopy (STS).
Interestingly, the observed bonding behavior turns out to be distinctly different on copper, silver and gold surfaces.
On Cu(001), TCNE forms tightly bound islands while loosely bound islands were identified on Au(111).\cite{Weg08,TCNEonCu}
In contrast to this, TCNE molecules on Ag(001) occur for low coverage as isolated monomers.\cite{Weg08}
In the latter case, a structural model has been worked out on the basis of the measured STM images.
An electron charge transfer from the silver substrate to the molecule has been inferred from a comparison
of the observed STS images with charge distributions of prominent molecular orbitals
resulting from calculations within density-functional theory for {\it gas-phase}\/ neutral
and negatively charged TCNE molecules.
Very recently, adsorption of ordered TCNE monolayers on Ag(001) has been 
investigated in a combined study\cite{TCNEonAgIETS} 
employing STM, inelastic electron tunneling spectroscopy and density-functional theory
to reveal the structural and vibrational properties of this system.
A theoretical investigation of {\it isolated}\/ TCNE adsorbed on Ag(001), however, is still missing.

In this paper, we investigate the structural and electronic properties of 
isolated TCNE molecules adsorbed on Ag(001) employing density-functional theory (DFT).
Our results explain interesting experimental findings\cite{Weg08} that cannot be deduced without
ambiguity from a calculation for a gas-phase TCNE molecule.
We show that the structural model proposed in Ref.~\onlinecite{Weg08} is confirmed by total energy calculations and 
that the prominent peak\cite{Weg08} at $-0.6$\,V in the experimental d$I$/d$U$ curves 
is in nice agreement with the calculated local density of states (LDOS).
Furthermore, a trench around the TCNE molecule is observed in the experimental STM images.\cite{Weg08}
Employing two-dimensional Fourier transform methods together with a spatial extrapolation of the wave function,
our calculated STS and STM images reveal these experimental findings.

Our particular issue concerns the basis-set expansion used in this work.
We employ Gaussian orbitals that are centered at the atom positions.
Such basis functions may offer advantages for systems with much vacuum or with strongly localized wave functions.
For the present focus of evaluation STM and STS images, however, localized orbitals suffer from the severe problem
that they have no amplitude at the position of the STM tip, thus yielding no image.
Here we show that additional orbitals slightly above the surface and the adsorbate molecule,
in combination with a spatial extrapolation technique,
allows to evaluate STM and STS images at any tip height in a well-defined and reliable way.
The present system of TCNE adsorbed on Ag(001) is an ideal test system for this approach
due to the vast amount of experimental data under various conditions.

This paper is organized in the following way. In Sec.~\ref{sec:Methods} the theoretical approach 
is described. After a short introduction to the subsystems (Sec.~\ref{ss:Adsorbate}) the structural  
and electronic properties of TCNE on Ag(001) are presented in Secs.~\ref{ss:Position} and \ref{ss:Electronic}, respectively.
Afterwards the resulting STM and STS images are discussed and compared to experiment (Sec.~\ref{ss:STM}).

\section{Methods}\label{sec:Methods}
The calculations are carried out within the local density approximation (LDA) of density-functional theory.
Norm-conserving pseudopotentials\cite{pp_ham} in Kleinman-Bylander form \cite{pp_kb} including 
non-linear core corrections \cite{pp_nlcc} and the exchange-correlation functional of Ceperley and Adler \cite{ElGas} 
in the parametrization of Perdew and Zunger\cite{lda_pz} are used in our work.
In addition, we carry out calculations based on the generalized gradient
approximation (GGA) in the formulation of Perdew and Wang.\cite{gga_pw} 
The wave functions are expanded in a basis of atom-centered Gaussian orbitals of s, p, d and s$^*$ symmetry
with several decay constants per atom. 
\footnote{We use decay constants of
$0.11$, $0.37$, $1.14$ and $4.31$ for carbon and of $0.31$, $0.83$, $2.20$
and $6.93$ for nitrogen. For the atoms of the topmost silver layer decay constants of 
$0.14$, $0.44$, $1.20$, $2.95$ and $7.81$ are employed. For all other silver atoms 
$0.22$, $0.77$ and $2.21$ are used. All decay constants are given in atomic units.}

The convergence of the basis has been carefully tested for the gas-phase molecule, for bulk Ag and for the clean Ag(001).
We have confirmed that the basis gives bond lengths and lattice constants in good agreement with other theoretical investigations and with experiment.

\subsection{Representation of the wave function outside the crystal}\label{ss:Wavefunction}
The wave function $\psi^\text{Gau}_{n, \vec{k}}$ inside the crystal and the molecule and in the vicinity
of the surface is well described by Gaussian orbitals.
In a STM or STS measurement, however, a tip probes the wave function several \AA ngstr\"om above the surface. 
For a reliable description of such experiments, it is essential to treat the decay of the wave function into the vacuum more accurately.
To this end, we define a reference plane at height $z_0$ above the surface and molecule such that the potential $V(\vec{r})$ is constant
($= V_\text{Vac}$) to reasonable accuracy above $z_0$. We then extrapolate the wave function above the reference plane
by the corresponding solutions of the Schr\"odinger equation in the vacuum.
Here we give a brief outline of this procedure that has been described in more detail in Ref.~\onlinecite{RohlfingSTM}.

In the following we use a periodic surface, i.e. all states can be classified by a two-dimensional wave vector
$\vec{k} = (k_x, k_y)$.
The wave function is expanded for $ z \ge z_0$ into a two-dimensional Fourier sum
\begin{align}
  \psi_{n, \vec{k}}(\vec{r}_\parallel, z) = \sum_{\vec{G}} a_{\vec{G}}(z) \exp(i(\vec{k} + \vec{G}) \vec{r}_\parallel)\label{eq:FTb}
\end{align}
with reciprocal lattice vectors $\vec{G} = (G_x, G_y)$.
The coefficients $a_{\vec{G}}(z)$ obey the Schr\"odinger equation for a particle in a constant potential ($V(\vec{r}) = V_\text{Vac}$), i.e.
\begin{align}
  - \frac{\hbar^2}{2m} \left( \frac{\partial^2}{\partial z^2} - (\vec{k} + \vec{G})^2 \right) 
a_{\vec{G}}(z) = (E_{n, \vec{k}} - V_\text{vac}) a_{\vec{G}}(z) . \label{eq:SGc}
\end{align}
They decay exponentially into the vacuum
\begin{align}
  a_{\vec{G}}(z) = a_{\vec{G}}(z_0) \exp(-\lambda_{\vec{G}}\,(z-z_0))\label{eq:ext}
\end{align}
with decay constants
\begin{align}
  \lambda_{\vec{G}} = \sqrt{(\vec{k} + \vec{G})^2 - \frac{2m}{\hbar^2} (E_{n, \vec{k}} - V_\text{vac})}. \label{eq:decc}
\end{align}
At the reference plane $z_0$, the coefficients are determined by
\begin{align}
  a_{\vec{G}}(z_0) = \sum_{\vec{r}_\parallel} \psi_{n, \vec{k}}^\text{Gau}(\vec{r}_\parallel,z_0) \exp(-i(\vec{k} + \vec{G}) \vec{r}_\parallel )\label{eq:FT}
\end{align}
from the wave function $\psi_{n, \vec{k}}^\text{Gau}$ evaluated in the Gaussian basis on a real-space mesh $\vec{r}_\parallel = (x,y)$.
Eventually, the wave function at the desired height $z$ is calculated from equation~(\ref{eq:FTb}).

This method works well for ``flat'' systems, i.e. for systems in which all outermost atoms form a rather uniform ``plane'' without holes, dips, steps, etc.
The ideal case would be a flat, clean surface or a dense monolayer of adsorbed molecules.
In such case, a reference plane (at height $z_0$) can be defined such that the distance to the nearest atoms underneath does not exceed $\sim 2$ - $3$\,\AA\
at any lateral position and the atom-centered localized basis yields a stable wave function representation everywhere on the reference plane.

For a single adsorbate, however, the reference plane at $z_0$ must be several \AA ngstr\"om above the molecule
(in order to guarantee $V(\vec{r}) = V_\text{Vac}$ in the entire reference plane).
In such a case, the distance between the {\it surface} and the reference plane is so large that the
contributions from the surface itself to 
$\psi_{n, \vec{k}}^\text{Gau}$ are not sufficiently described by the original Gaussian basis,
in particular at positions above the clean surface far away from the molecule.
For this reason, the basis has to be improved by {\it additional orbitals} in the empty space above the surface,
i.e. at positions other than atoms. A typical arrangement is shown in Fig.~\ref{fig:1_side}.
We find that four orbitals per site (one s orbital and three p orbitals) with a small decay constant (e.g. $0.12/\text{a}_\text{B}^2$ for a silver surface)
are sufficient to improve the description of the wave function significantly.
We have checked that the results are insensitive to the number, positions and decay constants of these additional orbitals.

\subsection{Calculation of STM and STS images}\label{ss:calcSTM}
For the calculation of STM and STS images we use the Tersoff-Hamann theory,\cite{TerHam, STM-Heinze}
which describes the tunneling current $I$ as a function of the applied voltage $U$ by the integration of the local density of states
\begin{align}
  I(U) \propto \text{sgn}(U) \int_{E_F}^{E_F +  e U} \sum_{n, \vec{k}}  \left\vert \hat{\mathcal{O}}~\psi_{n, \vec{k}}(\vec{r}) \right\vert^2 dE\label{eq:stm}
\end{align}
and their derivative with respect to the voltage by
\begin{align}
 \frac{\partial I}{\partial U} \propto \sum_{n, \vec{k}}  \left\vert \hat{\mathcal{O}}~\psi_{n, \vec{k}}(\vec{r}) \right\vert^2 \delta(E_{n, \vec{k}}, E_F + e U).\label{eq:sts}
\end{align}
The vector $\vec{r}$ denotes the position of the tip which is in most cases several \AA ngstr\"oms above the surface. 
The operator $\hat{\mathcal{O}}$ describes the type of the tip.\cite{Chen}
Since the experiments often cannot specify the details of their tips, we assume that the tip is of s-like nature, and thus set $\hat{\mathcal{O}} = 1$.
To account for the discrete energy spectrum resulting from the supercell approach and for the finite $\vec{k}$ point mesh employed in our work,
we include a spectral broadening of 0.2 eV for the calculation of STM and STS images.
In the following, we present {\it constant current}\/ images. To this end, the tunneling current is
calculated at various heights and the height of the tip for a constant current is determined by a linear interpolation.

\section{Results and Discussion}\label{sec:Results}
The system we investigate is a TCNE molecule adsorbed on a silver (001) surface.
The calculations are carried out in a supercell.
If not noted otherwise we use a slab with three layers containing $6\times6$ silver atoms per layer ($17.2\times17.2$\,\AA).
This leads to 118 atoms in a supercell.
The simulations of the STM images at large distances require a big lateral unit cell in the case of a single molecule on a surface.
For that purpose, we enlarge our cell to $8\times8$ silver atoms per layer, resulting in a $7440\times7440$ Hamilton matrix.
Special $\vec{k}$ points on a $4\times4\times1$ mesh according to Monkhorst and Pack\cite{MP76} are used for 
the evaluation of the $\vec{k}$ sums during the self-consistent cycle and for the computation of the STM and STS images. 

\subsection{Silver substrate and TCNE molecule}\label{ss:Adsorbate}
For fcc bulk silver we calculate a lattice constant (and bulk
modulus) of $a_\text{LDA} = 4.07\,$\AA\ ($B_\text{LDA} =
122\,\text{GPa}$) in LDA and $a_\text{GGA} = 4.21\,$\AA\ ($B_\text{GGA} =
86\,\text{GPa}$) in GGA. Comparing these results to the experimental values of
$4.07\,$\AA\ ($102\,\text{GPa}$)\cite{AllElAg} it is obvious that LDA gives
reasonable results whereas GGA overestimates the lattice constant distinctly. 
Similar deviations have been observed by other authors \cite{Haas} for this material.
For the clean surface we observe an inward relaxation of the topmost layer by
$-0.6$\% in LDA and $-0.4$\% in GGA, respectively. These values are slightly smaller than the results of other theoretical
calculations ($-1.3$\ldots$-2.0$\%) and comparable with experimental data ($0\pm1.5$\%).\cite{AgSurf}

TCNE is a small organic molecule consisting of a central ethylene double
bond surrounded by four cyano groups.
For the isolated neutral molecule in the gas phase we find bond lengths of
$d_{\text{C}=\text{C}}=1.37$\,\AA, $d_{\text{C}-\text{C}}=1.41$\,\AA\ and
$d_{\text{C}\equiv\text{N}}=1.15$\,\AA\ in LDA and
$d_{\text{C}=\text{C}}=1.38$\,\AA, $d_{\text{C}-\text{C}}=1.43$\,\AA\ and
$d_{\text{C}\equiv\text{N}}=1.16$\,\AA\ in GGA. All lengths differ 
less than $0.03\,$\AA\ from the experimental ones.\cite{TCNEexp}
We observe in LDA as well as in GGA angles of
$121^\circ$ between the ethylene bond and the cyano group and of 
$179^\circ$ within each cyano group. The corresponding experimental values \cite{TCNEexp} are 
$122^\circ$ and $178^\circ$, respectively.

Both exchange-correlation functionals (LDA and GGA) give reliable results for the isolated
TCNE molecule. Bulk silver, however, is much better described within 
LDA. Therefore, we employ the local density approximation for the
investigation of the combined system of TCNE on Ag(001). 

\subsection{Optimal Adsorption site}\label{ss:Position}
\begin{figure}[tb]%
  \includegraphics{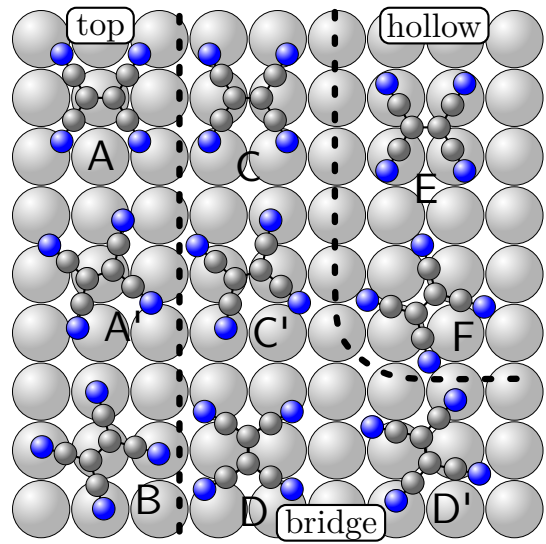}%
  \caption{(Color online) Different start positions for the structural optimization of TCNE
on Ag(001). Nitrogen, carbon and silver atoms of the topmost substrate layer are indicated by 
blue, dark gray and light gray circles, respectively.}\label{fig:pos}%
\end{figure}%

\begin{figure}[tb]%
  \includegraphics[scale=.59]{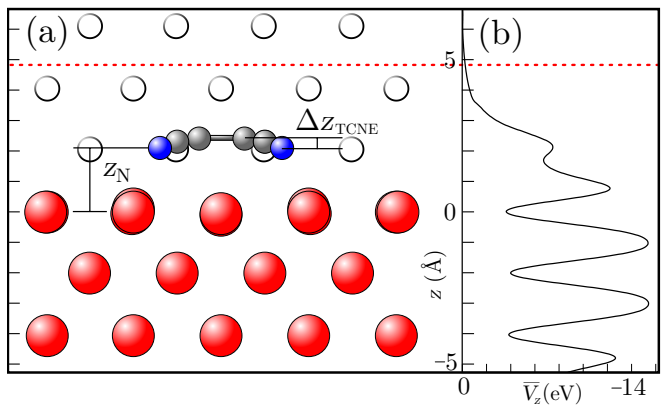}%
  \caption{(Color online) (a) Side view of optimized structure A and (b) potential 
$\bar{V}_z$ which follows from the Kohn-Sham potential by averaging over the lateral coordinates.
Carbon, nitrogen and silver atoms are indicated by blue, gray and red dots, respectively.
Empty circles denote centers of additional orbitals employed in the representation of the wave function.
The red dotted line shows the position of the reference plane in the height $z_0 = 4.9$\,\AA\ above the clean surface at $z = 0$\,\AA.
}\label{fig:1_side}%
\end{figure}

We carry out structural optimization for TCNE on Ag(001) employing various start geometries
which are shown in Figure~\ref{fig:pos}.
The configuration of an adsorbed molecule can be characterized by the position of the double bond
between the central carbon atoms and by the angle of orientation.
There are {\it top}, {\it hollow}\/ and {\it bottom}\/ positions that are investigated for
several angles (see Fig.~\ref{fig:pos}).
The molecule and the topmost layer are allowed to relax until the forces are less than $5\times 10^{-4}\,\text{Ry}/a_\text{B}$.
The structures {A'}, {C'} and {D'} rotate during the optimization 
procedure and eventually take the same final positions as {A}, {C} and {D}, respectively.
In all calculations we observe that the nitrogen atoms bind to the surface. As a result, the molecule
bends and is no longer planar like in vacuum. All resulting configurations show high molecular symmetry,
i.e. all four nitrogen atoms are found in equivalent positions and at equal height. As an example, we show 
 in Fig.~\ref{fig:1_side} the side view of the optimized structure for configuration {A}.
The silver atoms below the center of TCNE have moved by $-0.12$\,\AA\ toward the substrate
whereas those below the nitrogen have moved 
outwards by $0.09$\,\AA. To specify a configuration, we use the bond length $d_\text{Ag-N}$ between
a TCNE nitrogen atom and the nearest silver surface atom,
as well as the height $z_\text{N}$ which is the 
difference in the $z$ coordinates of the topmost silver layer of a clean surface and the nitrogen atom.
A further measure of the bending of the molecule is the intramolecular vertical distortion $\Delta z_\text{TCNE}$,
i.e. the height difference between the central carbon atoms and the nitrogen atoms.
The results for different adsorption positions are shown in 
Tab.~\ref{tab:struc_res} together with each corresponding adsorption energy.
\begin{table}[bt]%
 \centering
 \caption{Structural parameters for different final configurations of an adsorbed TCNE molecule in top (A, B), bridge (C, D) and hollow (E, F) positions.
Bond length $d_\text{Ag-N}$, height $z_\text{N}$ between the surface and TCNE, extension $\Delta z_\text{TCNE}$ of the molecule in $z$ direction
 and adsorption energies $E_\text{ads}$ (in eV) for different configurations. All lengths are given in \AA.}\label{tab:struc_res}
  \begin{ruledtabular}
  \begin{tabular}{cccccc}
    ~\#~  &$d_\text{Ag-N}$  & $z_{\text{N}}$  & $\Delta z_\text{TCNE}$  & $E_{\text{ads}}$ \\
   \hline
    {A} & $ 2.32$ & $2.12$ & $0.32$ & $2.72$ \\
    {B} & $ 2.30$ & $2.35$ & $0.43$ & $2.47$ \\[.1cm]
    {C} & $ 2.32$ & $2.24$ & $0.32$ & $2.52$ \\
    {D} & $ 2.32$ & $2.13$ & $0.69$ & $2.23$ \\[.1cm]
    {E} & $ 2.52$ & $2.34$ & $0.50$ & $1.89$ \\
    {F} & $ 2.77$ & $2.22$ & $0.22$ & $1.78$ \\
  \end{tabular}
  \end{ruledtabular}
\end{table}%
$E_\text{ads}$ is defined as the difference between the energies of the isolated parts and that of the interacting system, 
\begin{align}
  E_\text{ads} = - E_\text{Ag-TCNE} + E_\text{TCNE} + E_\text{Ag} - E^\text{BSSE},\nonumber
\end{align}
where a counterpoise correction $E^\text{BSSE}$ is included to compensate the basis set superposition error.
We calculate the correction for each investigated structure by the method suggested in Ref.~\onlinecite{BSSE},
in which the energy error due to the different basis sets is estimated.
The resulting energies $E^\text{BSSE}$ are nearly the same varying between $0.44$ and $0.49$\,eV for all structures.
For configurations \text{{A} - {D}} we observe bond lengths $d_\text{Ag-N}$ of about
$2.30$\,\AA\ while much larger bond lengths occur for configurations {E} and
{F}. The heights $z_\text{N}$ are in the range of $2.13$ to $2.35$\,\AA\
while the vertical distortion $\Delta z_\text{TCNE}$ varies between $0.22$ and $0.69$\,\AA.
For adsorption in bridge position {C}, we find the strongest distortion and bending of the molecule.

Configuration {A} with TCNE in the top position is energetically
most favorable. In this structure, four \text{Ag-N} bonds with a length of 2.32
\AA\ are formed with a moderate bending of the molecule. TCNE in the
bridge position {C} leads to an adsorption energy that is less favorable by
$0.2$\,eV. In configuration {D}, \text{Ag-N} bonds with a length of $2.32$\,\AA\ are
achieved at the expense of a stronger bending of TCNE which reduces the
adsorption energy. The hollow positions {E} and {F} are energetically very
unfavorable due to the larger \text{Ag-N} bond lengths in these configurations.
Adsorption of TCNE in the top position {A} is also indicated by
experiments in Ref. \onlinecite{Weg08} using STM with atomic resolution. In the following,
we give a more thorough discussion of the structural and
electronic properties of this particular configuration.

\begin{table}[bt]%
 \centering
 \caption{Structural parameters of TCNE adsorbed on Ag(001) in optimum top position {A},
as well as of gas-phase neutral TCNE and of single and double negatively charged TCNE$^{-}$ and TCNE$^{2-}$, respectively.
The lengths are given in \AA.
All calculations are carried out within the LDA.}\label{tab:struc_vgl}
  \begin{ruledtabular}
  \begin{tabular}{ccccc}
    ~  & TCNE & TCNE & TCNE$^-$ & TCNE$^{2-}$ \\
    ~  & on Ag(001) & gas phase & gas phase & gas phase \\
   \hline
    $d_{\text{C}=\text{C}}$      & $1.49$ & $1.37$ & $1.43$ & $1.49$ \\
    $d_{\text{C}-\text{C}}$      & $1.38$ & $1.41$ & $1.39$ & $1.38$ \\
    $d_{\text{C}\equiv\text{N}}$ & $1.17$ & $1.15$ & $1.16$ & $1.17$ \\
    $\Delta z_\text{TCNE}$       & $0.36$ & $0.00$ & $0.00$ & $0.00$ \\
  \end{tabular}
  \end{ruledtabular}
\end{table}%

In Tab.~\ref{tab:struc_vgl} we compare the structural parameters of TCNE on Ag(001)
with those of a gas-phase molecule. The \text{C=C} bonds between the central
C atoms, as well as the \text{N$\equiv$C} bonds are weakened in the adsorbed
state. This can be attributed to the adsorption-induced occupation of the
TCNE lowest unoccupied molecular orbital (LUMO), which is an antibonding orbital located at the two central C atoms
and the four N atoms of TCNE. The same changes of bond strength and bond length are found for negatively charged gas-phase TCNE,
as shown in Tab.~\ref{tab:struc_vgl}.
In particular for gas-phase TCNE$^{2-}$ we find
the same bond length as for adsorbed TCNE. A detailed discussion of the
adsorption-induced charge redistribution will be given in the next section.

\subsection{Electronic properties}\label{ss:Electronic}
\begin{figure}[tb]%
  \includegraphics[scale=.45]{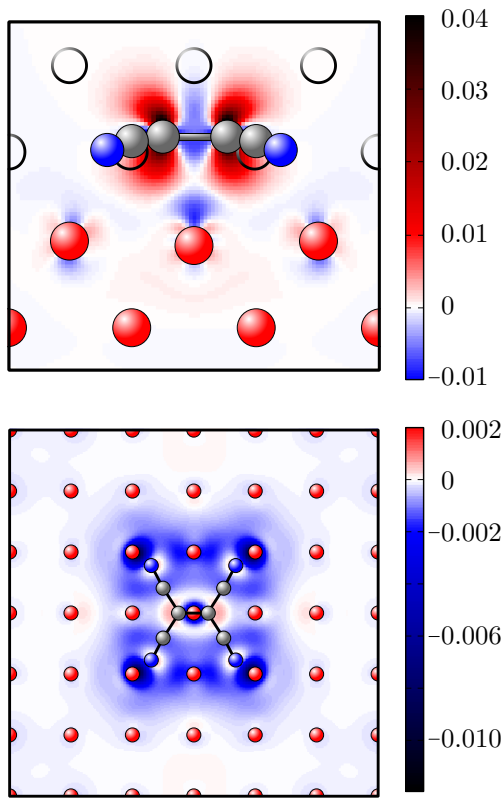}%
  \caption{(Color online) Electron density difference (in $e/a_\text{B}^3$)
for TCNE on Ag(001) in configuration A. 
The vertical cut (upper panel) is shown along the carbon double bond. 
The horizontal cut (lower panel) is plotted $0.8$\,\AA\ above the surface, i.e. below the TCNE molecule.
Nitrogen, carbon and silver atoms are indicated by blue, gray and red dots, respectively.
The open circles mark the position of additional orbitals used for the representation of the wave function.
The red color marks increased electron density, blue color decreased electron density. Note the different scales in both panels.
}\label{fig:1_chgdiff}%
\end{figure}%

To quantify the charge transfer we calculate the atomic charges using the Bader
analysis.\cite{Bader,BaderMeth} Summing over all atoms of the molecule, TCNE in configuration {A}
leads to a charge of $-1.3$\,e, which
means that more than one electron is transferred from the surface to the molecule.
Nearly the same values are obtained for the other investigated adsorption structures \text{{B} - {F}}.

\begin{figure*}[tb]%
  \includegraphics[scale=.75]{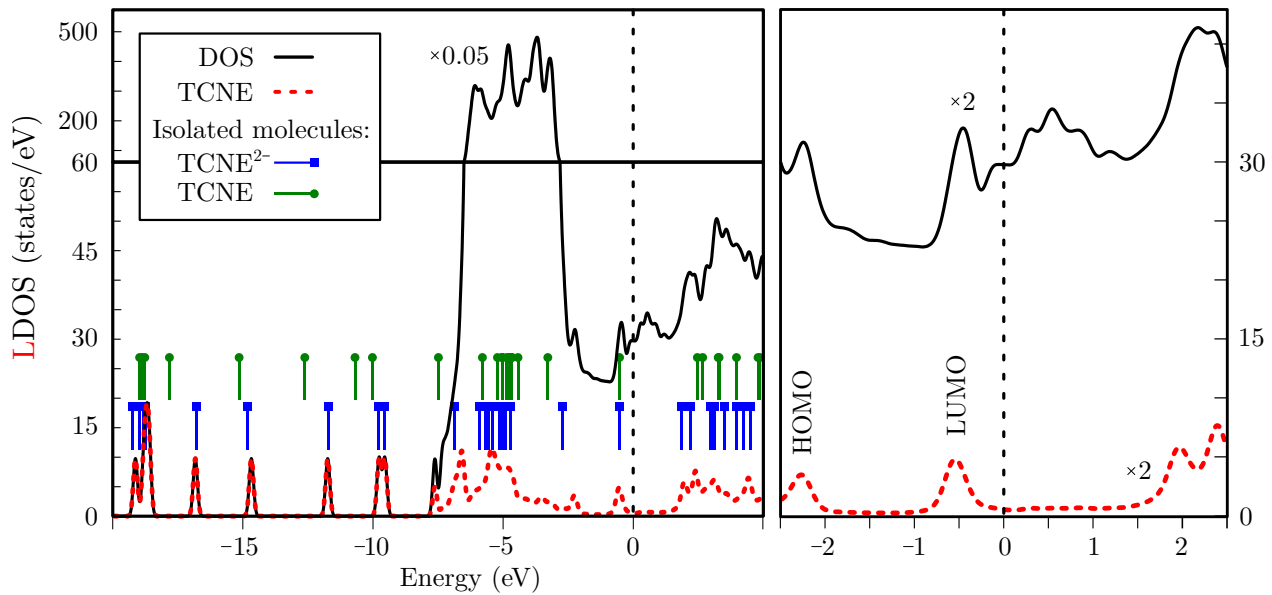}%
  \caption{(Color online) Total DOS (black line) and LDOS at TCNE (red dashed line) for configuration {A}.
The blue (green) marks indicate the discrete spectrum of gas-phase TCNE$^{2-}$ (TCNE).
The upper region of the DOS (above 60\,states/eV) is scaled down by a factor of 1/20. 
A section around the Fermi energy is shown scaled up by a factor of two in the right panel.
A $24\times24$ mesh of $\vec{k}$ points has been used in the calculations together with an energetic broadening of $0.08$\,eV.
}\label{fig:1_ldos}%
\end{figure*}%

To take a further look inside this process we compare in Fig.~\ref{fig:1_chgdiff}  the charge  
density difference between the adsorbate system and its isolated parts. 
A reduction of the electron density underneath the molecule and a large increase of the density near the
central carbon atoms is visible. 
The spatially localized increase of the density is given by the population of the LUMO of the TCNE molecule,
while the decrease of the density occurs in a more extended region of the Ag surface, mostly about 1\,\AA\ above the surface-atom nuclei.
Only a small density change takes place inside the substrate.
It is worth to notice that the use of the additional orbitals (see Sec.~\ref{ss:Wavefunction}) is mandatory for a
reliable description of the charge redistribution in this system.

Figure~\ref{fig:1_ldos} shows the density of states (DOS) for TCNE on Ag(001) in the optimal configuration {A},
calculated with a broadening of $0.08$\,eV. 
The local density of states  of the adsorbed TCNE molecule as resulting from
a Mulliken analysis\cite{Mulliken} of the wave function is given by the dashed red line.
For comparison we have indicated the energy levels of neutral gas-phase TCNE, as well as of negatively charged gas-phase TCNE$^{2-}$.
The energetic levels of both gas-phase molecules have been aligned in such a way that the states below $-8$\,eV fit best,
because the smallest adsorption induced changes are expected in this region.
At energies below $-8$\,eV the DOS coincides with the LDOS and reflects the energetic positions of the molecular levels of TCNE$^{2-}$.
We note in passing that the energy levels of TCNE$^-$ (not shown here) have positions similar to those of TCNE$^{2-}$.
Between $-8$ and $-2$\,eV there is a strong contribution of the Ag d orbitals to the DOS.
At higher energies, the DOS is dominated by Ag sp states.
In addition to the numerical broadening of $0.08$\,eV, we observe for energies above $-8$\,eV
a broadening of the molecular levels due to the interaction with the substrate.
The positions of the LUMO of gas-phase TCNE and of the highest occupied molecular orbital (HOMO) of gas-phase TCNE$^{2-}$
agree well with the position of the TCNE:Ag(001) LDOS peak at $-0.5$\,eV.
The LDOS corroborates that TCNE adsorbed on Ag(001) shows
close electronic resemblance with a gas-phase charged TCNE$^{2-}$ molecule, in addition to the 
close structural resemblance mentioned above.

As described above, we find a charge transfer of $-1.3$\,e but an occupation of nearly two electron in the former LUMO.
This indicates a charge transfer of $0.7$ electrons from various interacting orbitals of TCNE back to the substrate, which is known as back donation.

An enlarged section of the calculated electronic spectrum around the Fermi level is shown in the right panel of Fig.~\ref{fig:1_ldos}.
The two peaks in the LDOS of TCNE at $-2.3$\,eV and $-0.5$\,eV correspond to the HOMO and LUMO of TCNE, respectively.
Experimental d$I$/d$U$ curves in Ref.~\onlinecite{Weg08} show a clear peak at $-0.6$\,V
which was assigned to the LUMO of TCNE on the basis of a calculation for the gas-phase molecule.
Our results for TCNE on Ag(001) corroborate this interpretation of the STS data.

\begin{figure}[tb]%
  \includegraphics[scale=.38]{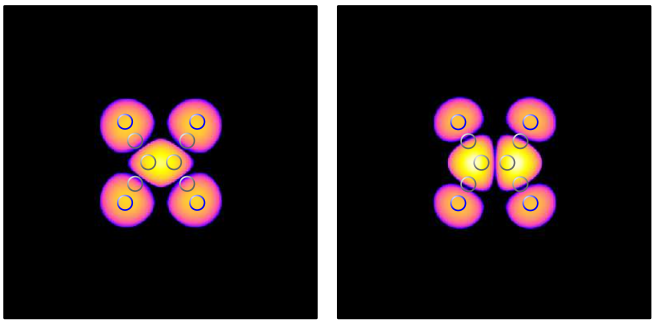}%
  \caption{(Color online) Density isosurface 
of HOMO (left panel) and LUMO (right panel)
for gas-phase TCNE at a density of $10^{-4}\,a_\text{B}^{-3}$.
}\label{fig:TCNE_rho2d}%
\end{figure}%

To set the stage for a discussion of the STS images of TCNE on Ag(001)
we first have a look at the charge densities of the HOMO and LUMO of an isolated TCNE molecule
that are shown in Fig.~\ref{fig:TCNE_rho2d}.
We note that the pictured isosurfaces are very close to those for the HOMO-1 and HOMO of gas-phase TCNE$^{2-}$, respectively.
The HOMO of TCNE is a $\pi$-bonding state with maximum density at the central carbon atoms.
In contrast to this, the LUMO shows a nodal plane at the center of the C=C bond,
which is the characteristic of an anti-bonding $\pi^*$ orbital.
Occupation of this orbital weakens the C=C bond leading to a bond length of $1.49$\,\AA\ at TCNE$^{2-}$ in comparison to $1.37$\,\AA\ at neutral TCNE.
For both states, about half of the electron density is located at the four nitrogen atoms.
In the optimal adsorption position for TCNE on Ag(001) the molecule is oriented in such a way that 
the N atoms come close to the substrate. Occupation of the $\pi^*$ orbital is accompanied by a
charge transfer from the substrate to the molecule as described above.

We have also calculated the LDOS for the other investigated structures.
The general features look similar but the peak positions are shifted towards higher energies.
In particular, the LUMO derived state is at $-0.3$\,eV for configurations {B}, {C} and {D} and at $-0.2$\,eV for {E} and {F}.
Summarizing, we find the best agreement between the peak at $-0.6$\,V in the experimental d$I$/d$U$
curves and the calculated LDOS for configuration {A}, which is also the energetically most favorable 
structure.

\subsection{STS and STM images}\label{ss:STM}
\begin{figure*}[tb]%
  \includegraphics[scale=.4]{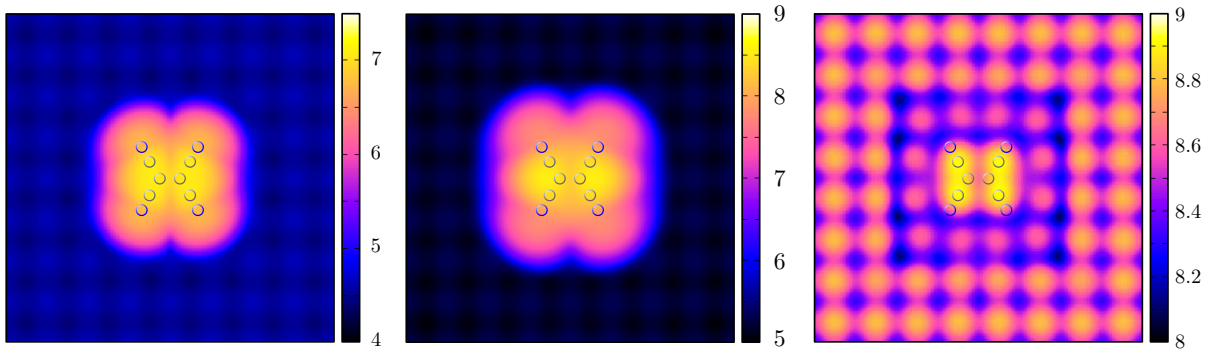}%
  \caption{(Color online) STM images at $-0.7$\,eV for configuration {A} calculated in a $8 \times 8$ unit cell without 
additional orbitals, by direct evaluation (left panel) and with spatial extrapolation (middle panel).
The result for direct evaluation with additional orbitals is shown in the right panel. The height is given in \AA.
All three images are physically questionable (see text).
}\label{fig:1_stm_comp}%
\end{figure*}

In experiment, topographical STM images provide information about the structure of 
adsorbates and surfaces combined with the electronic properties. 
Interestingly, STM images measured in Ref.~\onlinecite{Weg08} at a voltage of $-0.6$\,V show 
that the adsorbed TCNE molecule is surrounded by a trench.
In the trench the tip is $0.2$ - $0.3$\,\AA\ closer to the surface than on clean Ag(001).
Calculated STM images based on a direct evaluation of the wave function do not show these details
(see left panel of Fig.~\ref{fig:1_stm_comp}).
In the following we show that the refined treatment of the wave function outside the crystal as 
outlined in Sec.~\ref{ss:Wavefunction}
gives STM images with a trench in good agreement with experiment.

All calculated STM and STS images presented here are evaluated for a density
(i.e. right-hand side of Eq.~(\ref{eq:stm})) of 10$^{-9}\,a_\text{B}^{-3}$.
To include the contributions from the broad peak at $-0.5$\,eV in the LDOS, the STM images are  
computed for an energy of $-0.7$\,eV (i.e. all states between $-0.7$ and $0.0$\,eV contribute).
The STM image in the left panel of Fig.~\ref{fig:1_stm_comp}
is calculated with a basis of Gaussian orbitals localized at the atoms of the 
substrate and the molecule, respectively.
No additional orbitals above the surface (cf. open circles in Fig.~\ref{fig:1_side}) are considered here.
The tip is at a height of $5$ - $7$\,\AA\ above the surface for the desired density of 10$^{-9}\,a_\text{B}^{-3}$.
At this distance, the STM image is dominated by the contributions from the TCNE molecule
due to the strong decay of Gaussian orbitals, and the resulting image resembles that of the LUMO at TCNE.
In the next step, we extrapolate the wave function, starting from a reference plane at $z_0 = 4.9 $\,\AA\ as outlined in Sec.~\ref{ss:Wavefunction}.
The effective potential in the Kohn-Sham equation is almost zero for $z > z_0$ as can be inferred from Fig.~\ref{fig:1_side}(b).
Again, no additional Gaussian orbitals are included in the basis set.
The resulting density shown in the middle panel of Fig.~\ref{fig:1_stm_comp} decays exponentially for $z > z_0$
and reaches the value of 10$^{-9}\,a_\text{B}^{-3}$ at heights of $6$ - $9$\,\AA\ above the surface.
However, the experimentally observed trench is still missing since the contributions of the Gaussian orbitals
located at the Ag atoms have already decayed too strongly at the reference plane.

\begin{figure*}[htb]%
  \includegraphics[scale=.45]{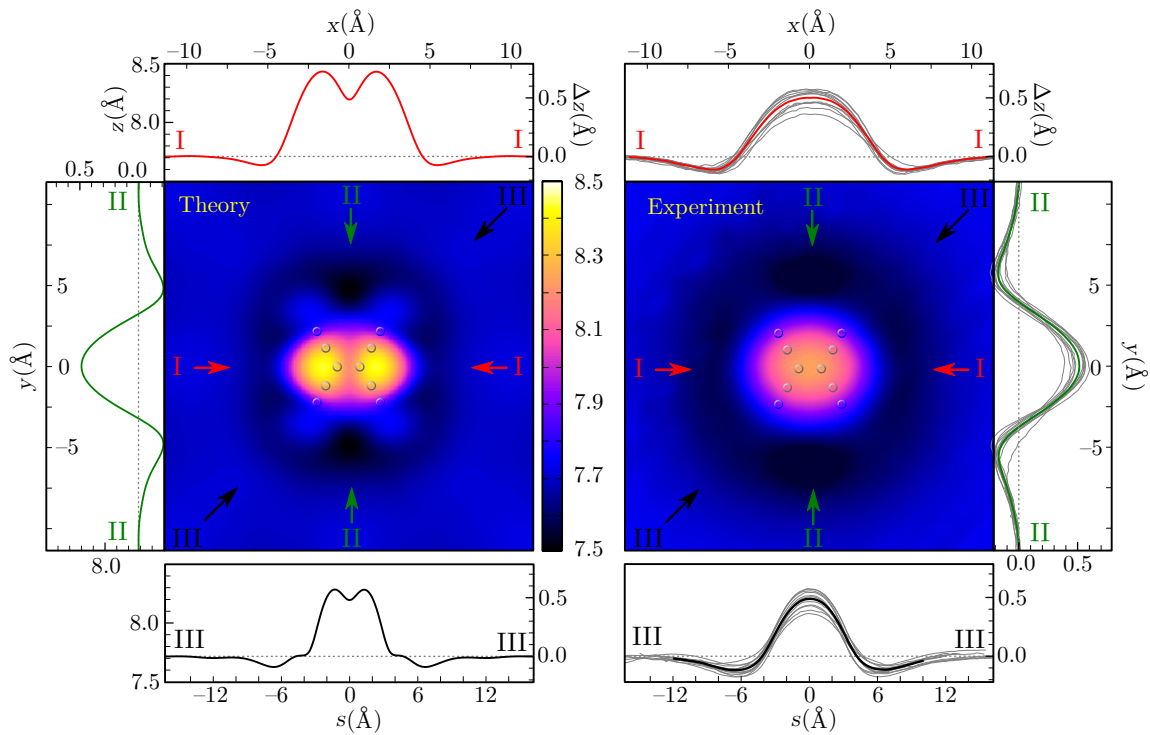}%
  \caption{(Color online) Left panel: STM image for configuration {A} at $-0.7$\,eV calculated in a 
$8 \times 8$ unit cell with additional orbitals and spatial extrapolation (constant-current image).
The height is given in \AA. Above, on the left and below this panel, scans along the lines I, II and III are given.
$s$ is the coordinate along line III.
Right panel: Experimental STM image at $U = -0.6$\,V.
The indicated positions of the atoms in the right panel are inferred from a comparison with the theoretical results.
The color range covers the same $\Delta z$ as in theory.
The line scans in gray show the results of 11 different samples. 
}\label{fig:1_stm}%
\end{figure*}%
\begin{figure}[htb]%
  \includegraphics[scale=.53]{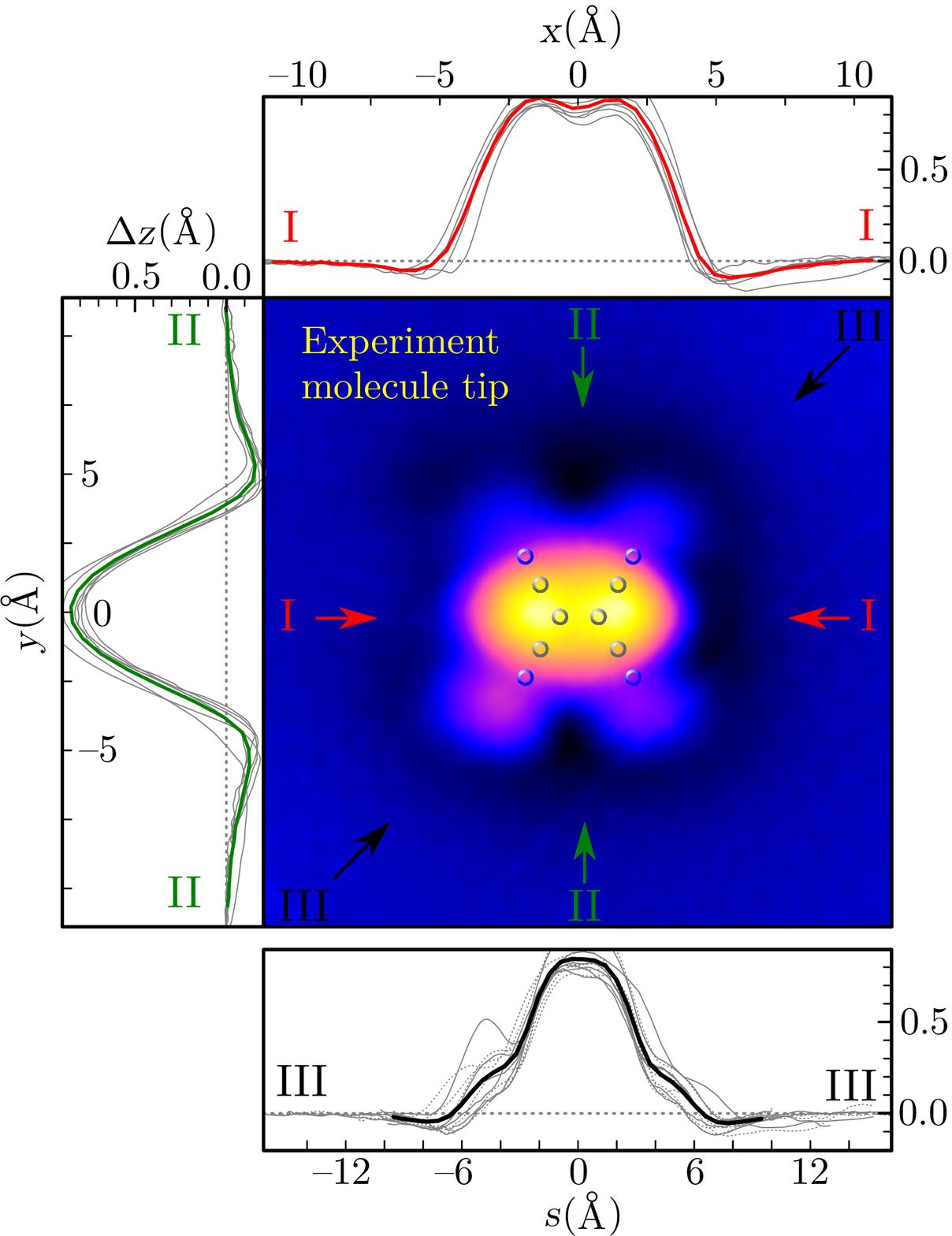}
  \caption{(Color online) Experimental STM image using a molecule tip.
The line scans in gray show the results of six different samples for three different molecule tips.
Because of the asymmetry of the images additional line scans (dotted) along the other diagonal are shown.
For more details see Fig.~\ref{fig:1_stm}. In comparison to Fig.~\ref{fig:1_stm} the range of the color scale is slightly increased to $1.1$\,\AA.
}\label{fig:1_stm-mol}%
\end{figure}%

The quality of the wave function $\psi^\text{Gau}_{n, \vec{k}}(\vec{r})$ can be improved in the region of the reference plane
by employing additional Gaussian orbitals in the empty space.
We use three layers of such orbitals located at those positions which atoms of further substrate layers would have (see Fig.~\ref{fig:1_side}).
Two of the layers reside between the silver substrate and the reference plane while a third one is above the reference plane.
Extrapolation of the wave function according to Sec.~\ref{ss:Wavefunction} leads to the 
STM image shown in the left panel of Fig.~\ref{fig:1_stm}.
The tip resides at $7.5$ - $8.5$\,\AA\ above the surface with the largest height above the carbon atoms.
In the region of the TCNE molecule, the image has an almost elliptical shape;
there are only small contributions from the N atoms to the tunneling current.
Most interestingly, the image does show a trench around the molecule, i.e. the tip resides closer to the surface than on the clean silver substrate.
This is further illustrated in Fig.~\ref{fig:1_stm} by three line scans indicated by I, II and III.
An experimental STM image with same scale is shown in the right panel of Fig.~\ref{fig:1_stm}.
The experimental line scans (in gray) are performed for 11 different samples, with their average marked by the colored thick line.
The theoretical results compare nicely with many features of the experimental STM image.
Both in theory and in experiment a trench around the molecule with its minima on line II can be identified.
In the calculation, the trench nearly vanishes at the positions of the cyano groups (line III).
In experiment, the trench is also significantly smaller along line III than along line I for most of the investigated samples.
The nodal line of the $\pi^*$ orbital is still visible in the calculated image (line I and III) while it is
probably smeared in experiment due to a larger distance of the tip (see the discussion of Fig.~\ref{fig:1_stm-side} below) and an oval shaped maximum occurs in the measured image.

The experimental STM resolution can be significantly improved by adsorbing a molecule at the tip.\cite{MolSTM}
In Fig.~\ref{fig:1_stm-mol} we present experimental data using such a molecule tip,
where we assume that a TCNE molecule is adsorbed.
The height corrugation of the image is increased to $1$\,\AA\ ($0.7$\,\AA\ with previous tips, cf. Fig.~\ref{fig:1_stm}) in comparison to $0.9$\,\AA\ in the calculation.
In contrast to the previously used tips, a minimum on the nodal line is indeed observed (line I).
In addition, the legs of TCNE are visible now (line III).
The height of these legs is slightly enlarged in comparison with the calculation,
while the minimum now resides at $s \approx \pm 7$\,\AA.
Summarizing, the agreement between calculation and experiment is good with a standard tip and even better when using a molecule tip.

The calculated STM image depends only weakly on the numerical details
(position of reference plane, decay constants and positions of additional orbitals).
For a moderate change of these values we obtain nearly the same results.
In contrast to this, a direct evaluation of the wave function far away from the surface
leads to artificial structures in the STM image, even when additional orbitals are employed,
as can be seen in the right panel of Fig.~\ref{fig:1_stm_comp}.
The contribution of the additional orbitals in the topmost layer dominates at distances $z > 8$\,\AA,
and the positions of the additional orbitals wrongly appear in the STM image.
Moreover, the results depend sensitively on the location and the decay constants of these orbitals.

\begin{figure}[tb]%
  \includegraphics[scale=.45]{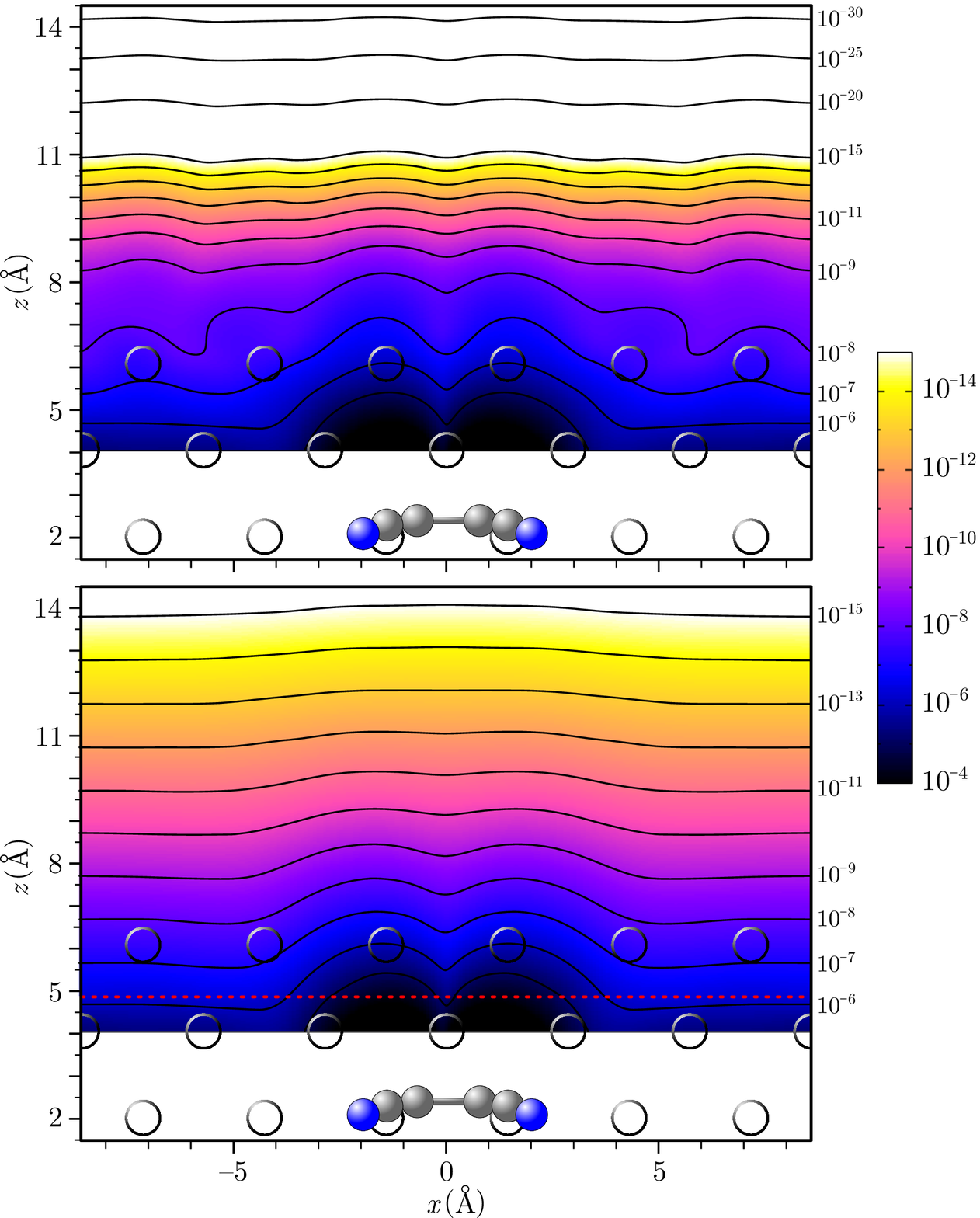}%
  \caption{(Color online) Vertical cut of density (Eq.~(\ref{eq:stm})) for configuration {A} with states between $-0.7$\,eV and $E_F$
calculated with direct evaluation (upper panel) and with spatial extrapolation (lower panel) of the wave function.
Additional orbitals have been employed in both cases.
The red dotted line marks the reference plane. Contours between $10^{-5}$ and $10^{-15}\,a_\text{B}^{-3}$ are indicated by black lines.
The horizontal axis is along line I (cf. Fig.~\ref{fig:1_stm}).%
}\label{fig:1_stm-side}%
\end{figure}%

The effect of these different calculational schemes is further illustrated in
Fig.~\ref{fig:1_stm-side} which shows (at various heights) the density for the states between $-0.7$\,eV and $E_F$,
calculated without (upper panel) and with (lower panel) spatial extrapolation of the wave function.
The horizontal axis is along line I (cf. Fig.~\ref{fig:1_stm}). %
The density contours with $\rho = 10^{-6}\,a_\text{B}^{-3}$ almost agree with each other for both computational methods.
At smaller densities, however, the outermost additional orbitals 
give rise to artificial variations of the contours in the direct evaluation (upper panel).
Such small densities occur at large distances from the surface where the contributions of 
the Gaussian orbitals located closer to the surface are strongly suppressed.
On the other hand, the wave functions are well represented at the reference plane 
and extrapolation then gives reasonable results far away from the surface.
The corrugation of the constant-density surface (i.e. the lateral variation of the height of constant density)
decreases with decreasing density.
Far away from the surface, the vertical movement of the tip becomes small.
At a density of $10^{-9}\,a_\text{B}^{-3}$, for example, the tip changes its vertical position by $0.7$\,\AA\ 
along the line scan II depicted in the left panel of Fig.~\ref{fig:1_stm}.
With respect to its position at a clean surface, the tip moves up by $0.5$\,\AA\ in the center 
of the TCNE molecule and comes closer to the surface by $0.2$\,\AA\ in the trench.
These movements are distinctly smaller than 
the distance of $2.4$\,\AA\ between the adsorbed TCNE molecule and the substrate.

The lower panel of Fig.~\ref{fig:1_stm-side} also shows a spatial widening of the molecular states as the tip moves up. 
For $\rho = 10^{-6}\,a_\text{B}^{-3}$, e.g., the two maxima of the constant-current line are $2.9$\,\AA\ apart from each other.
This corresponds to the distance of $2.5$\,\AA\ between the two main lobes of the TCNE LUMO as shown in Fig.~\ref{fig:TCNE_rho2d} (right panel).
On the other hand, for $\rho = 10^{-9}\,a_\text{B}^{-3}$ (which is reached when the tip has moved up by about $3$\,\AA),
the distance between the two maxima has increased to $3.4$\,\AA.
This clearly demonstrates that the lateral appearance of molecular features in STM images is height-dependent.
Atomic orbitals tend to ``move outwards'' and may no longer be observed directly above the atoms from which they originate.
We will come back to this issue when discussing the STS image.\\
Interestingly, the corrugation shows two maxima for densities $\rho$ larger than $10^{-13}\,a_\text{B}^{-3}$
(approximately 12\,\AA\ above the surface) while only \emph{one}\/ maximum is present
above the center of the molecule for smaller densities (i.e. at larger tip height).
In contrast to the measured STM image (right panel of Fig.~\ref{fig:1_stm}), in our calculation also the trench disappears at large tip height.
We suppose that the evaluation of the images at such large distances would require a larger lateral unit cell.

The other investigated adsorption configurations (\text{{B} - {F}}) show very similar STM images for $E = -0.7$\,eV. 
This is not surprising since the electronic structure in the energy interval from $-0.7$\,eV to $E_F$
is characterized for all configurations (\text{{A} - {F}}) by a state that resembles the LUMO of TCNE. 

A calculated STM image for unoccupied states up to $1$\,eV above $E_F$ also shows a faint trench around the TCNE molecule.
However, the distance between the bottom of the trench and the top of the molecule in this image is
considerably larger than in the measured one.\cite{Weg09}

\begin{figure}[tb]%
  \includegraphics[scale=.32]{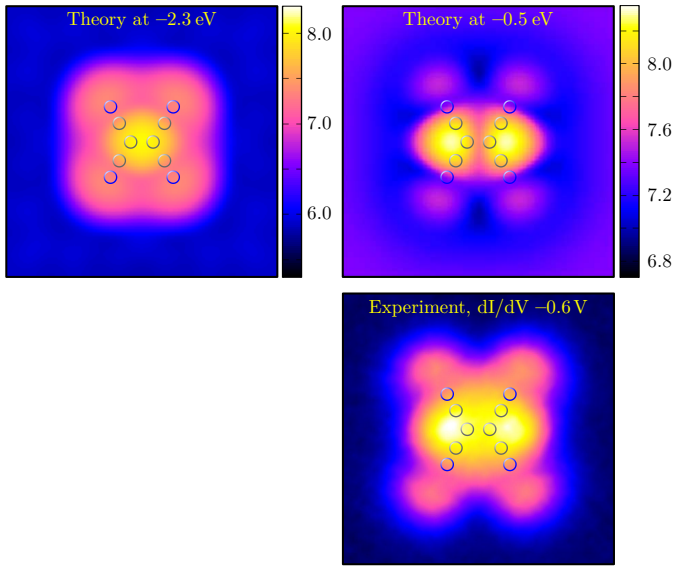}%
  \caption{(Color online) STS images for configuration {A} at $-2.3$\,eV (upper left panel) 
and $-0.5$\,eV (upper right panel) (calculated with spatial extrapolation and spectral broadening of $0.2$\,eV). 
The height is given in \AA. The empty circles 
denote the position of the atoms of TCNE. An experimental d$I$/d$U$ map at $-0.6$\,V is shown in the lower panel.
The indicated positions of the atoms in the lower panel are inferred from a comparison with the theoretical results.
}\label{fig:1_sts}%
\end{figure}%

Figure~\ref{fig:1_sts} presents calculated STS images for energies of $-2.3$\,eV and $-0.5$\,eV,
respectively, which correspond to the peak positions in the LDOS of TCNE as shown in the 
right panel of Fig.~\ref{fig:1_ldos}.
The STS image at $E = -2.3$\,eV with a main maximum at the \text{C=C} bond and local maxima at 
the four N atoms strongly resembles the charge density of the HOMO of TCNE. On the other hand, 
the image at $E = -0.5$\,eV is closely related to the charge density of the LUMO.
Comparing the upper right panel of Fig.~\ref{fig:1_sts} with the 
right panel of Fig.~\ref{fig:TCNE_rho2d} we see that 
the region of maximal values for TCNE on Ag(001) is more extended
than the atomic-orbital pattern closer to the atoms (as shown in Fig.~\ref{fig:TCNE_rho2d}).
In addition to this, the positions of the local maxima at the N atoms are shifted outwards.
This behavior corresponds to the general trend that molecular features appear laterally enlarged when probed at larger height,
as was already discussed for Fig.~\ref{fig:1_stm-side}.
We note that in the present case of TCNE, this trend is further supported by the distortion of the molecule.
The tilting of the C-N bonds by $9^\circ$ (relative to the flat gas-phase molecule) causes a corresponding
outward-tilting of the nitrogen p$_z$ orbitals (relative to the surface normal).
As a consequence, their contribution to the STM and STS images tends to move outwards,
as we observe in corresponding test calculations of flat and bent gas-phase molecules.
The calculated STS image at $E = -0.5$\,eV shows the same characteristic as the experimental 
d$I$/d$U$ map measured for this system at $U = -0.6$\,V (lower right panel of Fig.~\ref{fig:1_sts}):
an oval body intersected by a central line node and surrounded by four lobes.
Also in experiment, the lobes are not centered at the positions of the N atoms but they 
are further away from the molecule.
In summary, there is a nice agreement between theory and experiment.

All theoretical results presented above have been evaluated for a tip with s symmetry.
As mentioned in Sec.~\ref{ss:calcSTM}, tip orbitals with different symmetry, e.g. p$_z$ or d$_{3z^2 -r^2}$,
can be simulated by a corresponding operator $\hat{\mathcal{O}}$ in Eqs.~(\ref{eq:stm}, \ref{eq:sts}) (see Ref.~\onlinecite{Chen} for further discussion).
For the present adsorbate system, however, our investigations show that tips with p$_z$ or d$_{3z^2 -r^2}$ symmetry
give qualitatively very similar STM and STS images.

\section{Conclusions}
In summary, we have used density-functional theory to investigate the structural and
electronic properties of TCNE on the silver (001) surface in the limit of low coverage.
Our results support the structural model of Ref.~\onlinecite{Weg08} proposed on the basis of STM and STS measurements.
In the energetically most favorable structure, the C=C bond of TCNE is located above a silver atom
and the four cyano groups are bound to four neighboring silver surface atoms.
The bonding is accompanied by a transfer of about $1.3$ electrons from the substrate to the molecule due to occupation of the LUMO of TCNE.
This gives rise to a peak at $E = E_F - 0.5$\,eV in the LDOS of the adsorption system
in good agreement with a peak at $U = -0.6$\,V in the experimental d$I$/d$U$ curves.
Our results show that TCNE adsorbed on Ag(001) is
in close structural and electronic resemblance with a gas-phase charged TCNE$^{2-}$ molecule.
Employing a spatial extrapolation of the wave function into the vacuum region
we observe a faint trench in the calculated STM image at $E = E_F -0.7$\,eV 
in good agreement with the experimental results, especially with those obtained from a molecule-terminated tip.

\providecommand{\noopsort}[1]{}\providecommand{\singleletter}[1]{#1}%

\end{document}